\documentclass[10pt]{article}
\usepackage{amsmath,amsfonts,amssymb,a4,color,graphics}
\usepackage[latin1]{inputenc}
\newcommand{\R}{{\mathbb{R}}}

\newcommand{\N}{{\mathbb{N}}}

\def\ha{\frac{1}{2}}

\def\ra{\rightarrow}
\def\ga{\alpha}

\def\ge{\varepsilon}

\def\gf{\varphi}
\def\gg{\gamma}

\def\gl{\lambda}

\newtheorem{defi}{Definition}[section]

\newtheorem{lemm}{Lemma}[section]

\newtheorem{rem}{Remark}[section]
\newtheorem{coro}{Corollary}[section]
\newtheorem{theo}{Theorem}[section]

\newenvironment{demo}{\noindent {\it Proof.--}
      \begin{quotation}\noindent}{\end{quotation}\hfill$\square $}

\begin{document}

\title{Modes and quasi-modes on surfaces: \\
variation on an idea of Andrew Hassell }
\author{Yves  Colin de Verdi\`ere\footnote{Grenoble University, 
Institut Fourier,
 Unit{\'e} mixte
 de recherche CNRS-UJF 5582,
 BP 74, 38402-Saint Martin d'H\`eres Cedex (France);
{\color{blue} {\tt yves.colin-de-verdiere@ujf-grenoble.fr}}}}

\maketitle

\section{Introduction}

This paper is inspired from the nice idea   of
A. Hassell in 
\cite{Ha-Hi}.
From the classical paper of V. Arnol'd \cite{Ar}, we know that
quasi-modes are not always close to exact modes.
We will show that, for {\it almost all}
Riemannian metrics on closed surfaces 
with an  elliptic generic closed geodesic $\gg$,
  there exists exact  modes 
located on $\gg$. 
Similar problems in the integrable case are discussed 
in several papers of J. Toth and S. Zelditch (see 
\cite{T-Z2}).

\section{Quasi-modes associated to an elliptic
generic closed geodesic}\label{sec:quasi}

\subsection{Babich-Lazutkin and Ralston quasi-modes}
\begin{defi}\label{def:ell}
A periodic  geodesic $\gg $ on a Riemannian surface $(X,g)$ is said
to be {\rm elliptic generic} if the eigenvalues of the 
linearized Poincaré map of $\gg $ are of modulus $1$
and are not roots of the unity.
\end{defi}

\begin{theo}[Babich-Lazutkin \cite{Ba-La}, Ralston \cite{Ra,Ra2}]
\label{theo:ralston}
If $\gg $ is an  elliptic generic closed geodesic
of period $T>0$ on a closed Riemannian surface $(X,g)$, there exists a
sequence of 
{\rm quasi-modes}
$(u_m)_{m\in \N}$ of $L^2 (X,dx_g) $ norm equal to $1$
which satisfies 
\begin{itemize}
\item $\| (\Delta_g -\gl_m)u_m \|_{L^2(X,dx_g)}=O(m^{-\infty})$
\item There exists $\ga $ so that\footnote{$\ga $ is given
by $\ga = (m_1+\ha)\theta +p\pi $ where  $m_1 \in \N$ is a ``transverse''
quantum number, 
the linearized Poincaré map is a rotation
of angle $\theta $  
 ($0< \theta <2\pi $)  and $p=0$ or  $1$ is a ``Maslov index'' of $\gg$} 
 \[ \gl_m =\left(\frac{2\pi m +\ga }{T}\right) ^2 + O(1) \]
\item For any compact $K$ disjoint of $\gg$,
$\int_K |u_m|^2 =O(m^{-\infty})$.
\end{itemize}
\end{theo}
\begin{coro}
There exists a sub-sequence $(\mu_{j_m})_{m\in \N}$ of the spectrum
$(\mu_j)_{j\in \N} $ of the Laplace operator so that
$\mu_{j_m}=\gl_{m}+ O(m^{-\infty})$.
\end{coro}

\section{Modes and quasi-modes following Arnol'd }

Arnol'd \cite{Ar} has observed that, given a  quasi-mode
$(u_m)_{m\in \N}$,  there do not always exists a sequence
$(\gf_{j_m})_{m\in \N}$ of   exact modes close to the quasi-mode
$(u_m)_{m\in \N}$. 
His  example is given by a planar domain with a
symmetry of order $3$.

A simpler example is given  
by a symmetric double well:
let us give $V:\R \ra [0, +\infty[ $ a smooth even function
with
\begin{itemize}
\item $\lim_{x\ra \infty }V(x)=+\infty $
\item $V^{-1}(0)=\{ -a, a \}$ with $a>0$
\item $V(0)=b>0 $
\end{itemize}.
If $\hat{H}=-\hbar ^2 d_x^2 + V(x) $ is the semi-classical
Schrödinger operator, 
there exists quasi-modes located in  the  well $V:=
\{x~|~ x> 0~{\rm and ~}
V(x)< b\} $.
The exact  eigenfunctions are even or odd and hence are not localized 
in a single well.

The previous examples are in some sense {\it non generic}. They involve some 
symmetry of the operator.

\section{The main result}

\begin{theo} \label{theo:modes}
Let us give a closed Riemannian surface $(X,g_0)$
and a
 smooth non-zero function $f\geq 0 $.
Let us define the metric  $g_t:={\rm exp}(-tf)g_0 $.
Let us assume that there exists some  intervals 
$I_m =[\gl_m -l_m, \gl_m + l_m],~(m\in\N) $, 
{\bf independent of $t$},  so that, for any $t\in[0,1]$,
there exists at least one eigenvalue of $\Delta _t $
inside $I_m$. Assume that $\gl_m \ra + \infty $ and
$\sum _{m=1}^\infty l_m <\infty $.
Choose  a sequence $q_m\ra 0$ so that
$\sum _{m=1}^\infty  l_m/q_m <\infty $.

Then,  {\bf for almost all $t\in [0,1]$},
for any  sequence of exact modes 
 $\gf_{m} (t)$ of eigenvalues  $\mu_{m}(t)$  with $\mu_m(t) \in I_m$,
we have 
$ \int_X  f | \gf_{m}(t)|^2 dx_t  =o(q_{m} )$.

In particular, if $\Gamma ={\rm support }(f)$, 
$\gf_{m}(t) \ra 0 $ in $L^2_{\rm loc}(X\setminus \Gamma )$. 
\end{theo}

\begin{rem}
In applications, the interval $I_m$ is provided from
quasi-modes located in the support of $f$:
if $u_m $ is a quasi-mode for each values of $t$ 
with 
\[\|  (\Delta _t -\gl_m )u_m \| _{L^2(X,dx _0) }\leq 
C_m  \| u_m  \| _{L^2(X,dx _0) } \]
with $\gl_m $ independent of $t$, we can take 
$l_m =c C_m $ with $c$ large enough, depending only on bounds 
of $f$.

\end{rem}

\begin{rem}
The quasi-mode is only
used in order to
find a sequence of  intervalls $I_m$ 
which contains at least one eigenvalue
of $\Delta _t$  and is independent of $t$.
\end{rem}
\begin{rem}It works with $(u_m)$ the quasi-modes of Theorem
\ref{theo:ralston}  with 
 $\Gamma =\gg$ an elliptic generic closed geodesic
and $f$ flat  on $\gg$,
because the functions  $u_m$
satisfies an estimates
\[ u_m(x)=O\left(e^{-cd^2(x,\gg)/\sqrt{\gl_m }}\right)~.\]
 We can then take $q_m =O(m^{-\infty })$.  
 
We are unfortunately unable to prove
that the modes $\gf_{m}$ are close  
to linear combinations of the quasi-modes given in Theorem \ref{theo:ralston}
in the interval $I_{m}$.

The precise statement is
\begin{coro}
With the notations of Section  \ref{sec:quasi}, 
there exists a sequence $0< l_m=0(m^{-\infty })$
so that, for any $t\in [0,1]$, ${\rm Spectrum}(\Delta _t)\cap 
[\gl_m -l_m,\gl_m
+l_m ]\ne \emptyset  $
and a  subset $Z\subset [0,1]$ of measure $1$,
so that, for any sequence $\mu_{j_m}(t) \in [\gl_m -l_m,\gl_m
+l_m ] $
and for any $t\in Z$,
\[ \int _X f|\gf_{j_m}(t)|^2 dx_0 =0(m^{-\infty })~.\]

Moreover, for any compact $K\subset X$ with $K\cap \gg =\emptyset $
and for any $k\in \N $, 
we have
\[ \| \gf_{j_m}(t) \|_{C^k(K)}=0(m^{-\infty })~.\]
\end{coro}
\begin{demo} The first part is a direct application
of Theorem \ref{theo:modes}.

The second part comes from the Sobolev embeddings and the equations
 $\Delta_t ^N \gf_{j_m}(t)  =\mu_{j_m}(t) ^N  \gf_{j_m}(t) $
with $\mu_{j_m}(t)=0(m^2)$.
\end{demo}
\end{rem}

\begin{rem}
If we have only $l_m \ra 0 $, on can apply the previous result
by taking first a sub-sequence $m_k$ so that 
$\sum l_{m_k}< \infty $ and choosing then $q_{m_k} \ra 0 $.
This does not work with $l_m =O(1)$ as in the paper \cite{Ha-Hi}. 
\end{rem}

\section{Variation of the eigenvalues}
With $g_t=e^{-tf}g_0 $, we define $dx_t =e^{-tf}dx_0 $ the Riemannian
area of $g_t$ and $\Delta _t=e^{tf}\Delta _0 $ the
Laplace operator.
Let us denote by 
\[ \mu_1(t)=0 < \mu_2 (t)\leq \cdots \leq \mu_j(t) 
\leq \cdots \]  the eigenvalues of $\Delta _t$
 and by  $(\gf_j(t))_{j\in \N}$ an associated orthonormal eigenbasis.
\begin{lemm}
\begin{itemize}
\item $\mu_j(t) $ is a continuous  increasing function of $t$
\item $\mu_j(t)$ is piecewise analytic and, at any regular point, 
the $t$-derivative of $\mu_j(t)$ is given by:
\begin{equation} \label{equ:deriv}
 \dot{\mu_j}(t)=\mu_j (t)\int _X f \gf_j(t)^2 dx_t ~.\end{equation}
\end{itemize}
\end{lemm}
\begin{demo}
\begin{itemize}
\item The Rayleigh quotient $R_t(\gf)$ is given
by
\[ R_t(\gf)=\int _X \| d\gf \|^2_{g_0} dx_0 /
\int_X e^{-tf} \gf^2 dx_0 ~\]
which is an increasing function of $t$.
Applying the min-max characterization  of the eigenvalues,
we get their monotonicity. 
\item 
Because $\Delta _t$ is an analytic function of $t$,
we know that $\mu_j(t)$ is continuous and 
piecewise  smooth 
as well as $\gf_j(t)$.
We can then compute formally the derivative of the eigenfunction's equation
\[ e^{tf}\Delta_0 \gf_j(t) =\mu_j(t) \gf_j(t)~,\]
and get
\[f\Delta_t \gf_j(t)+\Delta_t \dot{\gf}_j(t)=
 \dot{\mu}_j (t)\gf_j (t)+ \mu_j(t) \dot{\gf}_j(t) ~,\]
and taking the $t$-scalar product with $\gf_j(t)$,
we get  Equation  (\ref{equ:deriv}).

 \end{itemize}

\end{demo}

\section{The proof}
The proof is an adaptation of the argument of  \cite{Ha-Hi}. 
Let us denote by $I_m=[\gl_m- l_m,\gl_m+ l_m]$.
From the Weyl law  and the monotonicity of the $\mu_j $'s,
we deduce the 
\begin{lemm}\label{lemm:weyl} For any $t\in [0,1]$,
$\# \{j~|~ \mu_j (t)\in I_m \} = O(\gl_m )$
uniformly in $t$.
\end{lemm}
In fact,
\[ \# \{j~|~ \mu_j (t)\in I_m \} \leq \# \{j~|~ \mu_j (t)\leq \gl_m +l_m
\}
\leq \# \{j~|~ \mu_j (0)
\leq \gl_m +l_m  \}~!\] 

We will also need the elementary
\begin{lemm}\label{lemm:estimate}
Let $F:[0,1] \ra \R $ be an increasing, continuous and 
piecewise $C^1$  function. Let us give a Borel set $Y\subset [0,1]$
and  $I $ a compact interval
of $\R$ so 
so that $F'(t)\geq m >0 $ for almost all $t\in K=F^{-1}(I)\cap Y $.
Then the Lebesgue measure $ |K|$  of $K$ satisfies
$ |K|\leq |I|/m $.
\end{lemm}
Let us denote 
by 
\[ Z:=\left\{ t \in  [0,1]~|~ \lim _{m\ra \infty} q_m^{-1}\left( \sup 
 _{\mu_j(t)\in I_m}\int_X f|\gf_j(t)|^2 dx_t \right) =0 \right\} ~,\]
($Z$ is well defined because there exists at least one $\mu_j (t)
\in I_m $ for each $m$) 
and $Y=[0,1]\setminus Z $.
Let us denote also, for $\ge >0$,   by
\[ Y_\ge^m :=\left\{ t ~|~\exists   j {\rm ~with~} \mu_j(t) \in I_m,
\int_X f| \gf_j(t)|^2 dx_t \geq \ge q_m \right\} ~.\]
Using Lemma \ref{lemm:weyl}, the monotonicity of $t\ra \mu_j(t)$ 
 and  the lower bound 
$\dot{\mu_j}(t)\geq \mu_j(t)  \ge q_m  $ in Lemma \ref{lemm:estimate}, we have
\[ |Y_\ge^m |\leq C \gl_m \frac{|I_m|}{ \ge q_m 
  (\gl_m -Cl_m )}=O\left( \frac{l_m}{\ge q_m} \right)~.\]

Let us give a sequence $\ge_m \ra 0 $,
then, for any $m_0$,
 $Y\subset \cup _{m\geq m_0} Y_{\ge_m}^m$.
But this implies that
$|Y|$  is arbitrarily close to $0$ by choosing $\ge_m $ so that 
$\sum_m  l_m/\ge_m q_m <\infty $. 
This proves that $|Z|=1$ and the Theorem.

\section{Null sets in Banach spaces}

It is not clear what is a set of measure $0$ in a infinite dimensional
 Banach space
because there is no  ``Lebesgue measure'' on it.
There are several definitions of sets of measure $0$
in a {\it separable Banach space} $B$. For Borel sets, it is shown
in \cite{Cso} that the notions of {\it cube} null sets
and {\it Gaussian} null sets coincide.

\begin{defi}
\begin{itemize}
\item
A {\rm cube measure} in $B$ is defined as the distribution of a random
variable $\sum_{i\in \N} t_i e_i$ where
${\bf t}=(t_i)\in [0,1]^\N $ with the Lebesgue measure
and the sequence $(e_i )_{i\in \N} $ span a dense 
subspace of  $B$ with
$\sum_{i\in \N} \| e_i \| <\infty $.
\item
A {\rm cube null set } is a Borel subset of $B$
which is of measure $0$ for every cube measure.
\item A {\rm Gaussian measure } on $B$ is a Borel  measure whose image
by any continuous linear form on $B$ is a (non-degenerate) Gaussian
measure on $R$ (i.e. 
of the form $dm=A.  {\rm exp}(-(x-a)^2/b ) dx $
with $A>0$).
\item A {\rm Gaussian null set} is a Borel set which
is of measure $0$ for every Gaussian measure. 
\end{itemize}
\end{defi}

It is proved in \cite{Cso} that cube null sets and Gaussian null
sets coincide in every separable Banach space.

We have the:
\begin{lemm}
Let $B$ be a separable Banach space and $C\subset B$
be a non empty  open cone. 
Let us give a Borel set $Z \subset B$
so that, for any $x\in B,~y\in C$,
\[ |\{ t~|~x+ty \in Z \}|=0 ~.\] 
Then $Z$ is a cube null  and Gaussian null set. 
\end{lemm}
\begin{demo}
Let us show that $Z$ is of measure $0$ for every cube measure
given from sequence $(e_i)_{i\in \N }$.
There exists $k\in \N $ and $(t_1,\cdots ,t_k ) \in [0,1]^k$
so that 
$e=\sum _{i=1}^k t_i e_i \in C $.
Let us rewrite the Lebesgue measure
on $[0,1]^k$ as
\begin{equation}\label{equ:fubini}
 \int_{[0,1]^k} f(t) dt = \int _X d\mu (d)\int _{d\cap [0,1]^k}f(t)
 ds 
\end{equation}
where $X$ is the set of lines parallel to $e$ cutting $[0,1]^k$
and $ds $ is the Lebesgue measure on the line $d$.
Let us denote  $t=(t',t'')\in  [0,1]^k \times [0,1]^\N$
and denote by \[ Z_{t''}:=\{ t' ~|~ x+ \sum _{i=1}^k t'_i e_i +
\sum_{j >k} t''_j e_j \in Z \} ~.\] 
Equation (\ref{equ:fubini}) shows that $Z_{t''}$ is of measure $0$.
We can then use Fubini Theorem  on $[0,1]^k \times [0,1]^{\N\setminus
\{ 1,\cdots, k\}} $ 
in order to finish the proof. 
\end{demo}

\section{From Theorem \ref{theo:modes} to almost all
metrics}

We will apply the previous result to the following situation
where $(X,g_0)$ is our  smooth closed surface and $\gg$
a closed geodesic; let us choose $N$
large (and even)  and define $B$ as follows:
\[ B=\{ f\in C^N(X,\R ) ~|~\forall \ga {\rm ~with~}
|\ga |\leq N ,~D^\ga f {~\rm vanishes ~on~}\gg \} \]
and $C $ the open cone of functions of $B$ which
satisfy 
\[ \exists c >0 {\rm ~such~that~}
f(x)\geq c d(x,\gg)^N \]
with $d$ the distance associated to $g_0$.

Then Theorem \ref{theo:modes} can be reformulated 
with {\it  almost all
metrics $e^f g_0 $ with $f \in B$}
instead of {\it  almost all $t\in[0,1]$} . Of course, we can only take
$l_m$ of the order of $ m^{-N'}$,
where $N'$ depends on  $N$.

\end{document}